\documentclass[11pt,showpacs,preprint,preprintnumbers,nofootinbib,superscriptaddress,amsmath,amssymb]{revtex4}
\usepackage{graphicx}
\usepackage{dcolumn}
\usepackage{bm}
\usepackage{amssymb}
\usepackage{amsmath}
\usepackage{epsfig}    
\usepackage{color}
\usepackage{slashed}
\usepackage{float}
\newcolumntype{I}{!{\vrule width 1.3pt}}
\begin{document} 
\title{Phenomenology of $E_6$-Inspired Leptophobic $Z'$ Boson at the LHC}

\author{Cheng-Wei Chiang}
\email{chengwei@ncu.edu.tw}
\affiliation{Department of Physics and Center for Mathematics and Theoretical Physics, 
National Central University, Chungli, Taiwan 32001, ROC}
\affiliation{Institute of Physics, Academia Sinica, Taipei, Taiwan 11529, ROC}
\affiliation{Physics Division, National Center for Theoretical Sciences, Hsinchu, Taiwan 30013, ROC}
\author{Takaaki Nomura}
\email{nomura@ncu.edu.tw}
\affiliation{Department of Physics, National Cheng Kung University, 1, Ta-Hsueh Road, Tainan, Taiwan 70101, ROC}
\author{Kei Yagyu}
\email{keiyagyu@ncu.edu.tw}
\affiliation{Department of Physics and Center for Mathematics and Theoretical Physics,
National Central University, Chungli, Taiwan 32001, ROC}

\begin{abstract}

We study collider phenomenology of a leptophobic $Z'$ boson existing in eight scenarios of the $E_6$ grand unified theory, differing in particle embeddings.
We first review the current bound on the $Z'$ mass $m_{Z'}$ based upon the LHC data of $pp\to t\bar{t}$ process at 8 TeV collisions with an integrated luminosity of 19.6 fb$^{-1}$.  Most scenarios have a lower bound of about 1 TeV.
However, this constraint does not apply to the case where $m_{Z'} < 2 m_t$, and other methods need to be employed for this lower mass regime.
Using existing UA2 constraints and dijet data at the LHC, 
we find that only one of the eight scenarios is excluded at 95\% confidence level.  No bound can be obtained from $Wjj$ and $Zjj$ measurements.
We propose to use the photon associated production of the $Z'$ boson that subsequently decays into a pair of bottom quarks, $pp\to Z'\gamma \to b\bar{b}\gamma$, at the LHC to explore the constraints in the lower mass regime. 
We compute the expected signal significance as a function of $m_{Z'}$ using detailed simulations of signal and irreducible background events.  
We find constraints for two more scenarios using the 8-TeV data and taking appropriate kinematical cuts.  
We also show the discovery reach for each scenario at the 14-TeV LHC machine.

\end{abstract}


\maketitle
\newpage

\section{introduction }

The operation of the 7- and 8-TeV runs of the CERN Large Hadron Collider (LHC) has provided us with quite important information about electroweak symmetry breaking; namely, 
the discovery of a standard model (SM)-like Higgs boson~\cite{Higgs} with a mass of about 126 GeV.
This fact becomes a strong guidance for us to consider various models beyond the SM. 
Moreover, null results of any other new particles so far impose lower bounds on their masses and/or new physics scales. 
It is of great interest to discuss what kind of signals from new physics can be expected at the upcoming 13-and 14-TeV runs, while taking into account the data collected in the 8-TeV run. 

An extra $U(1)$ gauge symmetry is often introduced based on various motivations in physics beyond the SM,  
resulting in an additional massive neutral gauge boson usually called the $Z'$ boson. 
For example, there are usually additional $U(1)$ gauge symmetries in grand unified theories (GUT's) such as the $E_6$ model~\cite{e6}. 
Besides, the discrete $Z_2$ symmetry required for stabilizing dark matter candidates can naturally emerge from a local $U(1)$ gauge group~\cite{residual_Z2,radseesaw,e6s4}. 
An extra $U(1)$ has also been employed in supersymmetric models~\cite{UMSSM} (so-called UMSSM) 
to facilitate a strong first order electroweak phase transition, as required to realize successful electroweak baryogenesis~\cite{ewbg}. 
Properties of such $Z'$ bosons strongly depend on the origin of the corresponding $U(1)$ symmetry in models. 
Therefore, phenomenological studies of $Z'$ bosons are essential to distinguish such new physics models~\cite{Langacker}. 

Searches for $Z'$ bosons have been performed mainly using the dilepton events at the LHC. 
If the couplings of the $Z'$ boson with fermions are the same as those of the $Z$ boson (the so-called sequential $Z'$ case),
the lower mass limit has been found to be 2.86 TeV (1.90 TeV) at the 95\% confidence level (CL) from 
collisions at 8 TeV with an integrated luminosity of 19.5 fb$^{-1}$ by using 
$e^+e^-$ and $\mu^+\mu^-$~\cite{ATLAS_Zp_dilepton} ($\tau^+\tau^-$~\cite{ATLAS_Zp_tau}) events. 

However, such searches become ineffective when the $Z'$ boson does not couple to the leptons. 
In this paper, we focus on the study of leptophobic $Z'$ bosons derived from different scenarios of the $E_6$ GUT.

In the $E_6$ model~\cite{babu}, there is a kinetic mixing between the hypercharge $U(1)_Y$ group and the extra $U(1)$'s after GUT breaking. 
As a result, the $Z'$ charge of each fermion is a linear combination of these $U(1)$ charges, involving two free parameters. 
They can be chosen so that the $Z'$ charges for the left-handed and right-handed charged leptons are zero, rendering the leptophobia nature. 

Phenomenological studies of the leptophobic $Z'$ boson had been done in Refs.~\cite{lpzp_tev,lpzp_lhc_dm,Buckley,lpzp_tevlhc,wjj,lpzp_precision}, with collider signals for the $Z'$ searched for at the Tevatron~\cite{lpzp_tev,lpzp_tevlhc} and the LHC~\cite{lpzp_tevlhc}. 
LHC collider signatures of a leptophobic $Z'$ boson that couples to a dark matter candidate had been studied in Refs.~\cite{lpzp_lhc_dm}. 
In Refs.~\cite{Buckley,wjj}, a leptophobic $Z'$ boson with the mass of about 150 GeV was proposed to explain the excess in the $Wjj$ events observed by the Tevatron CDF Collaboration.  
Effects of the leptophobic $Z'$ to the $e^+e^-\to q\bar{q}$ process due to the $Z$-$Z'$ mixing had been analyzed in Ref.~\cite{lpzp_precision}. 

In this paper, we discuss all possible scenarios with a leptophobic $Z'$ boson in the $E_6$ GUT model, differing in particle embeddings~\cite{London}.  
First, we consider the bound on the $Z'$ mass according to current data of collider experiments. 
In most scenarios, the $Z'$ can be excluded up to about $\mathcal{O}$(1) TeV by the $pp\to Z'\to t\bar{t}$ data at the LHC. 
However, this method does not apply when the $Z'$ mass is below the threshold for decaying into a pair of top quarks. 
We also take into account dijet data at the LHC and at the UA2, deriving respectively
a lower bound of about 500 GeV and 250 GeV on the $Z'$ mass only in one of the scenarios. 
In addition, although the $Wjj$ and $Zjj$ processes have been measured at the LHC, no bound can be obtained currently because of a small $Z'$ contribution to the cross sections compared to the experimental error bar. 
Therefore, we propose a promising channel, the photon associated production of $Z'$, at the LHC to search for 
the leptophobic $Z'$ boson with a mass smaller than $2m_t$.

We further focus on the bottom quark pair decay mode of $Z'$, $pp\to Z' \gamma \to b\bar{b} \gamma$, for the advantage of using double b-tagging to reduce the background events.
With a detailed simulation of signal and background events, we obtain the result of signal significance as a function of the $Z'$ mass. 
In addition, we further estimate the integrated luminosity required for a 5-sigma discovery for the 14-TeV LHC.

The structure of this paper is organized as follows.  
We review the interaction Lagrangian for the leptophobic $Z'$ boson in Section~\ref{sec:model}, where the decay and production of the $Z'$ are also discussed. 
The current bounds on the $Z'$ mass from various experiments are reviewed in Section~\ref{sec:bounds}. 
In Section~\ref{sec:simulation}, we propose to use the $pp\to Z'\gamma \to b\bar{b}\gamma$ process to search for a light $Z'$ boson.  A detailed simulation is presented to show what constraints we could have using the current data and the prospect of detecting such a particle at the 14-TeV LHC.
Our findings are summarized in Section~\ref{sec:conclusions}.
A brief review of the different leptophobic scenarios in the $E_6$ model is given in the Appendix.

\section{Leptophobic $Z'$ boson \label{sec:model}}

The interactions of the leptophobic $Z'$ boson with SM quarks are given by 
\begin{align}
\mathcal{L}&=\sum_{q=u,d}g_{Z'}\bar{q}\gamma^\mu (v_q-\gamma_5 a_q)qZ_\mu' ~, 
\label{Zp_int}
\end{align}
where $u$ and $d$ represent the up- and down-type quarks, respectively.  For simplicity, we assume no or at least negligible flavor-changing couplings.  The vector coupling coefficient $v_q$ and the axial-vector coupling coefficient $a_q$ are related to the $Z'$ charge $\bar{Q}_f$ of the quark $q$ by
\begin{align}
v_q=\frac{\bar{Q}_Q}{2} \left( 1+ \frac{\bar{Q}_q}{\bar{Q}_Q} \right) ~,\quad 
a_q=\frac{\bar{Q}_Q}{2} \left( 1- \frac{\bar{Q}_q}{\bar{Q}_Q} \right) ~,\quad 
(\text{for }q=u,d) ~.\label{va}
\end{align}
The appendix briefly reviews the scenarios in the $E_6$ GUT model that realize leptophobia for the $Z'$ boson, along with the corresponding $Z'$ charges. 
We note here that the value of the gauge coupling constant $g_{Z'}$ at the TeV scale can be predicted according to renormalization group running from the GUT scale, which depends on the details of matter contents and unification scale.  As in Ref.~\cite{Rizzo}, we adopt for definiteness $g_{Z'}=\sqrt{5/3}g'$, where $g'$ is the hypercharge coupling, for phenomenological analyses. 
In our paper, the non-SM fermions such as $h$ listed in Table~IV in the Appendix are assumed to be so heavy that their effects on the $Z'$ phenomenology can be safely neglected.

\begin{figure}[t]
\begin{center}
\includegraphics[width=80mm]{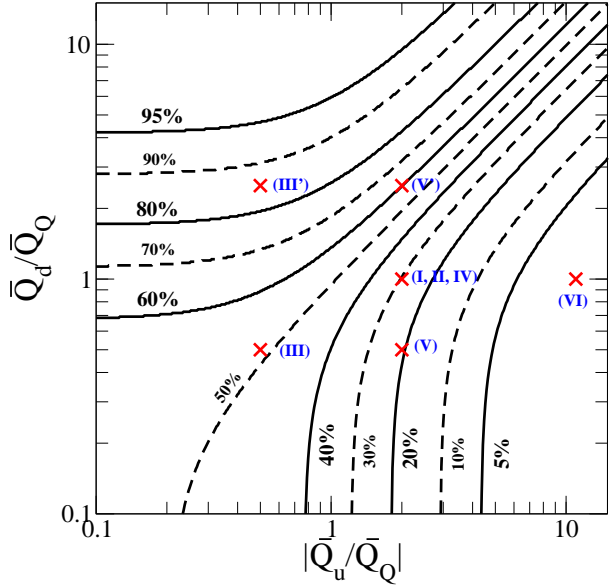}
\caption{
Contour plot of the total branching fraction for the $Z'$ decaying into the down-type quarks on the $|\bar{Q}_u/\bar{Q}_Q|$ and $\bar{Q}_d/\bar{Q}_Q$ plane in the case of $m_{Z'}=1$ TeV.  Predictions for the scenarios defined in Table~\ref{charges} in the appendix are indicated by the red crosses. 
}
\label{contour}
\end{center}
\end{figure}

The decay rate of $Z'$ into a quark pair is
\begin{align} 
\Gamma(Z'\to q\bar{q}) &=g_{Z'}^2\frac{m_{Z'}}{4\pi}[v_q^2(1+2x_q)+a_q^2(1-4x_q)]\sqrt{1-4x_q}\notag\\
& = g_{Z'}^2\frac{m_{Z'}}{4\pi}\frac{\bar{Q}_Q^2}{2}
\left[ 1+ \left(\frac{\bar{Q}_q}{\bar{Q}_Q} \right)^2-x_q \left(1- \frac{\bar{Q}_q}{\bar{Q}_Q} \right)^2 \right]\sqrt{1-4x_q}, 
\quad \text{with}~~x_q= \frac{m_q^2}{m_{Z'}^2}. 
\label{decay_rate}
\end{align}
Apart from $x_q$ that depends on the $Z'$ mass, the terms inside the square brackets involve just the two $Z'$ charge ratios, $\bar{Q}_u/\bar{Q}_Q$ and $\bar{Q}_d/\bar{Q}_Q$.  
Among the eight scenarios given in Table~\ref{charges}, 
Scenario-I and Scenario-IV have the same $Z'$ charges, and Scenario-II
have the same charge ratios as them, so that the decay branching fractions are the same in these three scenarios.
However, the total width and the production cross section for $Z'$ can be different between 
Scenario-I or Scenario-IV and Scenario-II as they are affected by the overall $\bar{Q}_Q$ factor.

In Fig.~\ref{contour}, we show the contour plot of the total branching fraction for the $Z'$ decaying into a pair of down-type quarks, 
$\sum_{q=d,s,b}\mathcal{B}(Z'\to q\bar{q})$, on the $|\bar{Q}_u/\bar{Q}_Q|$ and $\bar{Q}_d/\bar{Q}_Q$ plane, taking $m_{Z'}=1$ TeV as an example.  
Here we note that in the calculations of total width and branching fractions, the $Z'$ boson is assumed to decay into only SM quarks.
The reason we use the absolute value of $\bar{Q}_u/\bar{Q}_Q$ as the horizontal axis is simply because $\bar{Q}_u$ is always negative among the scenarios.
The predictions for the various leptophobic scenarios are indicated by red crosses. 
The maximum (about 85\%) and the minimum (less than 5\%) are realized in Scenario-III' and Scenario-VI, respectively. 
We note that varying the $Z'$ mass will cause shifts in the contours as a result of the $x_q$ dependence in Eq.~(\ref{decay_rate}).
The general tendency is that the total down-quark branching fraction becomes smaller for larger $Z'$ mass, as reflected in Table~\ref{br_tab}, where the branching fraction of each mode and the total width are computed for $m_{Z'}=500$, 1000 and 1500 GeV. 

In the following discussions, we will concentrate on the six scenarios that present distinct branching fraction patterns of the $Z'$, {\it i.e.}, Scenario-I, Scenario-III, Scenario-III', Scenario-V, Scenario-V', and Scenario-VI.

\begin{table}[t]
\begin{center}
\begin{tabular}{l|c|c|c|c} \hline \hline
Scenario & $u\bar{u}$ [\%] & $t \bar{t}$ [\%] & $d \bar{d}$ [\%]&width [GeV]\\ \hline
Scenario-I &&&& 3.2,~7.2,~11.1  \\ \cline{1-1}\cline{5-5}
Scenario-II &27.6,~24.7,~24.2&11.8,~20.8,~22.5&11.0,~9.9,~9.7&0.32,~0.72,~1.1  \\ \cline{1-1}\cline{5-5}
Scenario-IV &&&&3.2,~7.2,~11.1  \\ \hline
Scenario-III & 18.4,~17.1,~16.9& 7.9,~14.4,~15.7 & 18.4,~17.1,~16.9 &0.73,~1.6,~2.4 \\ \hline
Scenario-III' & 5.0,~4.9,~4.9 & 2.2,~4.2,~4.6 & 29.3,~28.7,~28.5 &2.7,~5.4,~8.2 \\ \hline
Scenario-V & 31.5,~27.8,~27.2 & 13.5,~23.5,~25.3 &7.9,~7.0,~6.8 &1.7,~3.9,~5.9    \\ \hline
Scenario-V' &14.8,~13.9,~13.7 &6.3,~11.7,~12.8&21.4,~20.2,~19.9 &3.6,~7.7,~11.7     \\ \hline
Scenario-VI & 37.9,~34.0,~33.3 & 22.3,~30.4,~31.7 &0.6,~0.6,~0.5&5.7,~12.8,~19.6   \\ \hline\hline 
\end{tabular}
\caption{Branching fractions and total width of $Z'$ in the scenarios listed in Table~\ref{charges}. 
The column for $u\bar{u}$ ($d\bar{d}$) displays the branching fractions for $Z'\to u\bar{u}$ and $Z'\to c\bar{c}$ ($Z'\to d\bar{d}$, $Z'\to s\bar{s}$ and $Z'\to b\bar{b}$). 
The three numbers in each entry are predicted for $m_{Z'}=500$ GeV, 1000 GeV and 1500 GeV, respectively. 
}
\label{br_tab}
\end{center}
\end{table}

\begin{figure}[t]
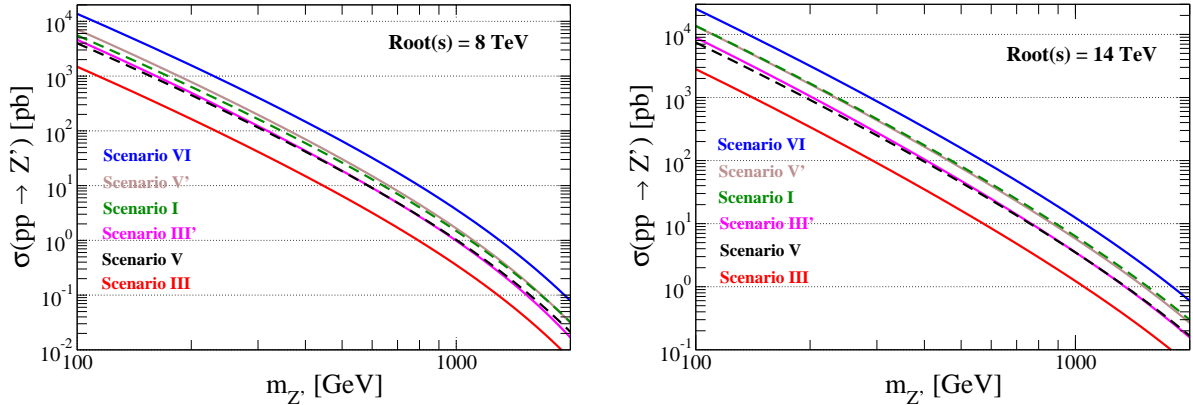

\begin{center}
\includegraphics[width=75mm]{cross_Zp_8TeV.eps} \hspace{5mm}
\includegraphics[width=75mm]{cross_Zp_14TeV.eps} 
\end{center}
\caption{
Production cross sections of the $pp\to Z'$ process as a function of $m_{Z'}$ for the 
collision energy of 8 TeV (left) and 14 TeV (right) in the six scenarios.
}
\label{cross_Zp}
\end{figure}

\begin{figure}[!t]
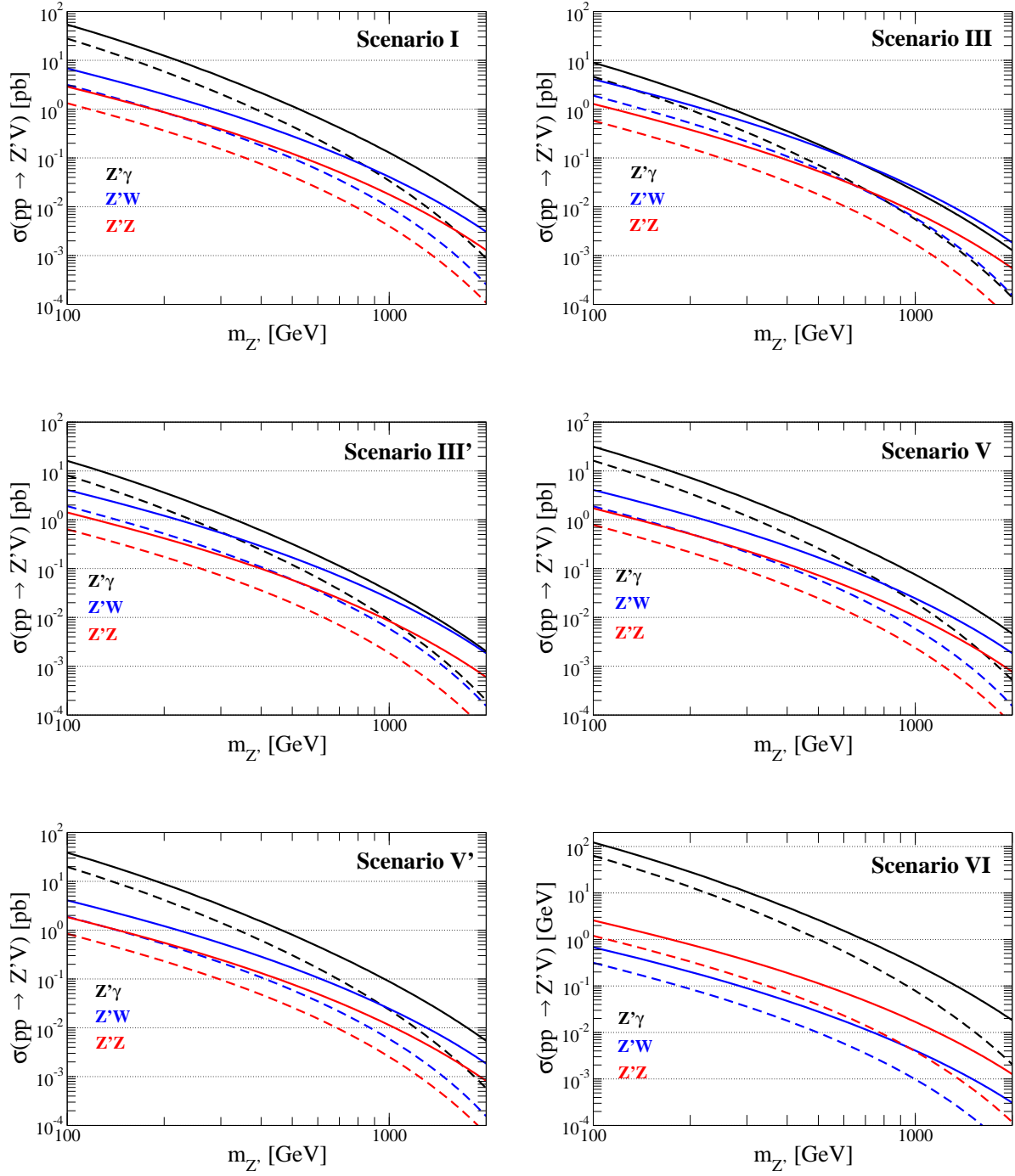
 
\begin{center}
\includegraphics[width=75mm]{cross_ZpV_I.eps} \hspace{5mm}
\includegraphics[width=75mm]{cross_ZpV_III.eps} \\
\vspace{9mm}
\includegraphics[width=75mm]{cross_ZpV_IIIp.eps} \hspace{5mm}
\includegraphics[width=75mm]{cross_ZpV_V.eps} \\
\vspace{9mm}
\includegraphics[width=75mm]{cross_ZpV_Vp.eps} \hspace{5mm}
\includegraphics[width=75mm]{cross_ZpV_VI.eps} 
\caption{
Cross sections of the $pp \to  Z' V$ ($V=\gamma,Z$ and $W^\pm$) process as a function of $m_{Z'}$ for the 
collision energy of 8 TeV (dashed curves) and 14 TeV (solid curves).
For the $pp\to Z'\gamma$ process, we impose $p_T(\gamma)>10$ GeV to avoid the collinear singularity. 
}
\label{cross_ZpV}
\end{center}
\end{figure}

Dominant production mechanisms for the $Z'$ at the LHC are the $s$-channel $pp\to Z'$ process and the $t$-channel $pp\to Z'V$ ($V=\gamma,Z$ and $W^\pm$) process of associated production.
We calculate the production cross sections for these processes and those in the subsequent analyses with the help of {\tt CalcHEP}~\cite{CalcHEP} package and using {\tt CTEQ6L} for the parton distribution functions (PDF's). 
In Fig.~\ref{cross_Zp}, the $s$-channel production cross section is shown as a function of $m_{Z'}$ for 
the collision energy of 8 TeV (left panel) and 14 TeV (right panel). 
The biggest (smallest) cross section in the whole mass range is given by Scenario-VI (Scenario-III), because of the larger (smaller) $\bar{Q}_u$ charge for the up-type quarks. 

In Fig.~\ref{cross_ZpV}, the associated production cross sections for the $pp\to Z'\gamma$, $pp\to Z'Z$ and 
$pp\to Z'W$ processes for each of the scenarios are shown as a function of $m_{Z'}$ by the black, red and blue curves, respectively, also for the collision energy of 8 TeV (dashed curves) and 14 TeV (solid curves). 
For the $pp\to Z'W$ process, the $W^+$ and $W^-$ contributions are summed over. 
We impose the $p_T(\gamma)>10$ GeV cut for the $pp\to Z'\gamma$ process to avoid collinear singularity of the produced photon, where $p_T(\gamma)$ denotes the transverse momentum for the photon. 
In most cases, the cross sections are generally ranked in the order of $Z'\gamma$, $Z'W$ and $Z'Z$ except for Scenario-VI.  In the large $m_{Z'}$ region, the produced vector boson tends to get smaller transverse mass, so that the production cross section of $Z'\gamma$ reduces faster than those of $Z'Z$ and $Z'W$, as a result of the $p_T(\gamma)$ cut.

\section{Constraints on the $Z'$ mass by current data \label{sec:bounds}}

In this section, we discuss various constraints on the $Z'$ mass in the six scenarios. 
We consider the current data on $pp \to  Z' \to t \bar{t}$, dijet and 
$W/Z$ plus dijet processes
at the LHC, as well as the $Z'{\bar q}q$ couplings extracted from UA2 experiment.

\subsection{The $pp\to t\bar{t}$ process}

\begin{figure}[t]
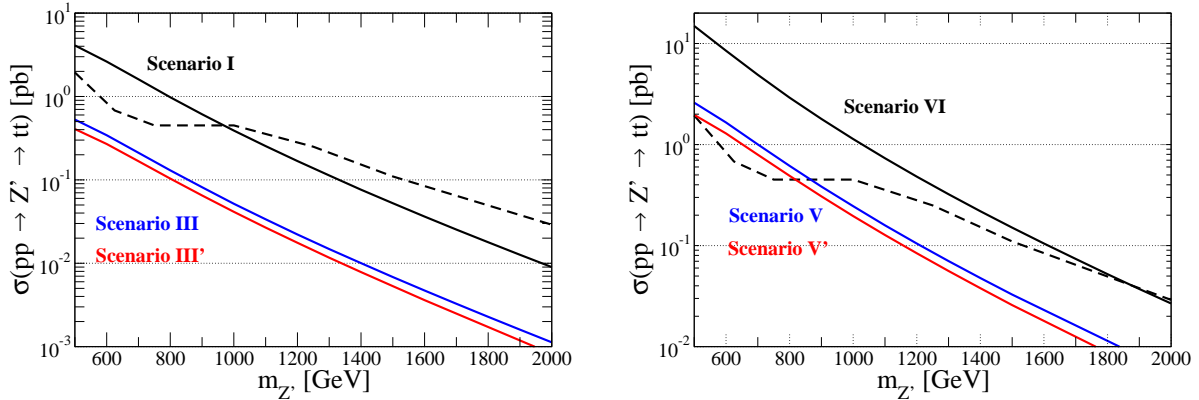
 
\begin{center}
\includegraphics[width=75mm]{bound_ttbar_1.eps} \hspace{5mm}
\includegraphics[width=75mm]{bound_ttbar_2.eps}
\caption{
Cross section of the $pp \to  Z' \to t \bar{t}$ process as a function of $m_{Z'}$ for the collision energy of 8 TeV. 
The left (right) panel shows the results in Scenario-I, -III and -III' (-V, -V' and -VI). 
The dashed curve is the observed limit at the 95 \% CL from the LHC data~\cite{CMStt}.  }
\label{ttConst} 
\end{center}
\end{figure}

The CMS group reported the search for production of heavy resonances decaying into $t \bar{t}$ pairs in Ref.~\cite{CMStt}. 
They analyzed the events with one muon or electron and at least two jets in the final state using the data corresponding to an integrated luminosity of 19.6 fb$^{-1}$ at 8 TeV.
Since no excess in events is observed, they provide an upper limit on the cross section of producing $t \bar{t}$ resonances at 95\% CL
as a function of invariant mass of $t \bar{t}$ pair $M_{t \bar{t}}$.
Comparing this limit with the cross section of $pp\to Z'\to t\bar{t}$ in our scenarios, 
we can obtain a constraint on $m_{Z'}$ by identifying $M_{t \bar{t}}$ as $m_{Z'}$. 

In Fig.~\ref{ttConst}, we show the cross section of the $pp\to Z'\to t\bar{t}$ process as a function of $m_{Z'}$ 
in Scenarios-I, -III and -III' (left panel) and in Scenarios-V, -V' and -VI (right panel). 
The experimental upper limit for the cross section is indicated by the dashed curve. 
In this calculation, the narrow width assumption is employed; {\it i.e.}, $\Gamma_{Z'}/m_{Z'}=1.2\%$ with $\Gamma_{Z'}$ denoting the total width of $Z'$, as used in Ref.~\cite{CMStt}.
In Scenarios-I, -V, -V' and -IV, the lower bounds on $m_{Z'}$ are extracted to be about 1000 GeV, 850 GeV, 800 GeV and 1900 GeV, respectively. 
On the other hand, no constraint is obtained from this process in Scenario-III and Scenario-III'.

\subsection{The dijet process}

\begin{figure}[t] 
\begin{center}
\includegraphics[width=150mm]{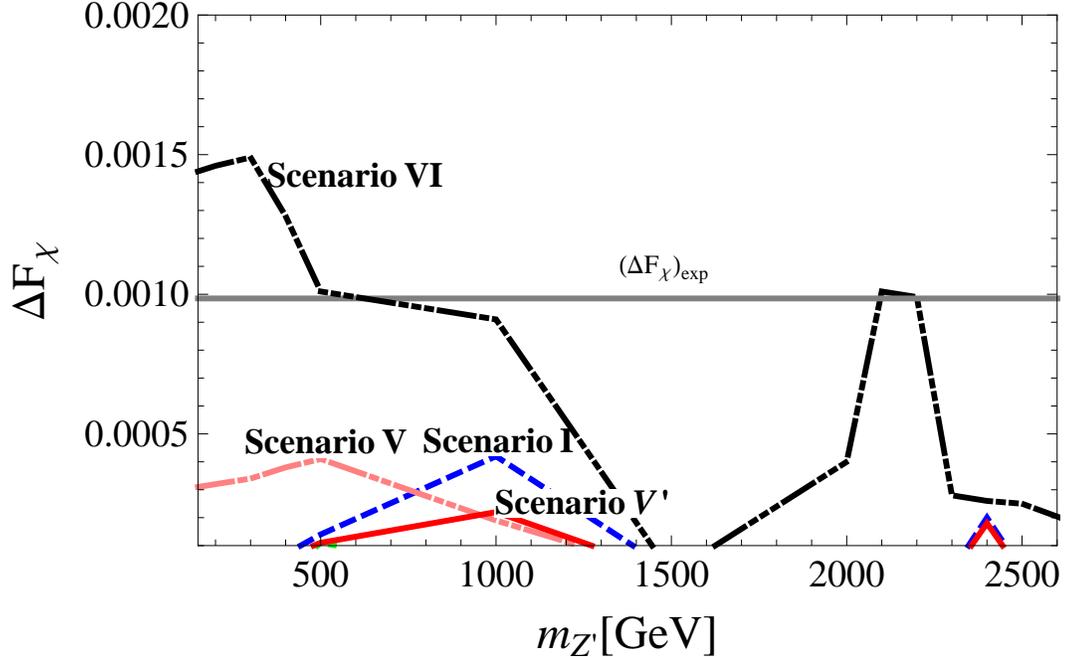}
\caption{
Values of $\Delta F_\chi$, defined in the text, in the different scenarios.
The horizontal line corresponds to the upper limit for $\Delta F_\chi$ at 95\% CL. 
\label{F_Chi} 
}
\end{center}
\end{figure}

Dijet events at hadron colliders are useful for the detection and property analysis of a leptophobic $Z'$ boson as one can readily measure a peak in the dijet invariant mass and study their angular distribution. 
Using the dijet resonant events at the LHC, Ref.~\cite{Dobrescu} put constraints on the mass and coupling constant for the leptophobic $Z'$ boson.  In Ref.~\cite{pomarol}, it had been shown that the angular distribution of the dijet events with a high invariant mass could be used to probe the mass scales associated with different dimension-6 operators by through effective Lagrangian approach. 
Particularly of interest to us is that the dijet process can receive contributions from four-quark interactions mediated by a $Z'$ boson when
the $Z'$ mass is taken to be much larger than the typical jet momentum. 
Such a new interaction term can modify the dijet distribution from that of the QCD prediction.
Thus, using the experimental dijet events, one can constrain the $Z'$ mass provided the coupling constant is fixed.  

The dijet measurement had been done at the LHC, offering various distributions~\cite{dijet_LHC1,dijet_LHC2}. 
In order to compare the dijet events including the $Z'$ mediation with the experimental data, 
we define $F_\chi$ value~\cite{dijet_LHC1} as the ratio of two numbers of events in different regions of $\chi$  as 
\begin{align}
F_\chi (m_{jj}^{\text{cut}})\equiv \frac{N(\chi < 3.32,m_{jj}^{\text{cut}})}{N(\chi < 30.0,m_{jj}^{\text{cut}})}, 
\end{align}
where $\chi\equiv\exp(|y_1-y_2|)$ with $y_{1,2}$ being the jet rapidities, and $m_{jj}^{\text{cut}}$ denotes an 
invariant mass cut for the dijet system. 
Ref.~\cite{dijet_LHC2} provides the dijet event data with $m_{jj}^{\text{cut}}$ as $2~\text{TeV}<m_{jj}<2.6~\text{TeV}$ based on the 4.8 fb$^{-1}$ data at 7 TeV.  The central value of $F_\chi$ is extracted to be about 0.0848. 

We directly calculate the deviation in the value of $F_\chi$ from the SM prediction with the help of {\tt MADGRAPH}~\cite{Ref:MG} and {\tt CTEQ6L}, instead of working with the effective Lagrangian.  This is because the mass region considered here is not large enough compared to the dijet invariant mass.  
The deviation can be expressed as 
\begin{align}
\Delta F_\chi\equiv \text{MIN}| F_\chi^{Z'}(1 \pm 0.02) - F_\chi^{\text{SM}}(1 \pm 0.02) |, 
\end{align}
where $F_\chi^{Z'}$ and $F_\chi^{\text{SM}}$ are respectively the values of $F_\chi(m_{jj}^{\text{cut}})$ 
in the six scenarios of the leptophobic $Z'$ and in the SM. 
In numerical calculations, there is about 2\% uncertainty in the cross section.  We therefore insert the factor $(1\pm 0.02)$ and pick the minimum on the right-hand side. 
When the SM prediction is assumed to be the same as the experimental central value, 
one can set a 95\% CL upper limit on $\Delta F_\chi$ by requiring  
$\Delta F_\chi< 1.96/\sqrt{N_{\exp}}$, where $N_{\exp}~(=28462)$ is the number of events measured at the LHC~\cite{dijet_LHC2}. 

In Fig.~\ref{F_Chi}, we show values of $\Delta F_\chi$ as a function of $m_{Z'}$ for the six scenarios. 
The 95\% CL upper limit for $\Delta F_\chi$ is indicated by the horizontal line. 
As shown in this figure, 
a $Z'$ with mass smaller than about 500 GeV is excluded only in Scenario-VI. 
On the other hand, no significant deviation in $F_\chi$ can be found in all the other scenarios. 
The peak at around 2 TeV for Scenario-VI is due to our choice of the dijet invariant mass cut, $2$ TeV $\leq m_{jj} \leq 2.6$ TeV.

\subsection{The $W/Z$ plus dijet events}

The $Wb\bar{b}$ and $Zb\bar{b}$ events had been measured at the LHC.
The measured cross sections of $pp \to Wb\bar{b}\to\mu \nu b\bar{b}$ and $pp \to Zb\bar{b}\to \ell^+ \ell^-b\bar{b}$ ($\ell$ is $e$ or $\mu$) processes
are given as $0.53 \pm 0.05~({\rm stat.}) \pm 0.09~({\rm syst.}) \pm 0.06~({\rm th.}) \pm 0.05~({\rm lum.})$ pb~\cite{Chatrchyan:2013uza}
and $0.37 \pm 0.01 ({\rm stat.}) \pm 0.07 ({\rm syst.}) $ pb~\cite{Chatrchyan:2014dha}, respectively, 
with 7 TeV and $5.0$ fb$^{-1}$. 
According to Ref.~\cite{Chatrchyan:2013uza}, the cross section of the $Wb\bar{b}$ event has been obtained by taking 
the kinematical cuts $p_T > 25$ GeV and $|\eta| < 2.1$ for muon, and $p_T > 25$ GeV and $|\eta| < 2.4$ for b-tagged jets.
On the other hand, 
the following cuts are imposed to obtain the cross section of the $Zb\bar{b}$ event:
$p_T > 20$ GeV,  $|\eta| < 2.4$ and $76 < M_{\ell \ell} < 106$ GeV for charged leptons, and $p_T > 25$ GeV and $|\eta| < 2.1$ for b-tagged jets~\cite{Chatrchyan:2014dha}.

We calculate the cross sections of the $pp \to Wb\bar{b}\to\mu \nu b\bar{b}$ and $pp \to Zb\bar{b}\to \ell^+ \ell^-b\bar{b}$ processes
in the SM and in Scenario-I with $m_{Z'} =100$ GeV using {\tt MADGRAPH}~\cite{Ref:MG}.
We find that the deviation in these cross sections are less than $0.01$ pb and $0.002$ pb for $Wb\bar{b}$ and $Zb\bar{b}$ processes, respectively, 
which are much smaller than the above-mentioned experimental errors.
Similar results can be obtained in all the other scenarios and heavier $Z'$ masses.  Thus, we cannot obtain any useful constraint on $m_{Z'}$ using the current data of these processes.

The CMS group also analyzed $Wjj$ events with the invariant mass of the dijet system $M_{jj}$ to be about 150 GeV. 
This process can receive contributions from the leptophobic $Z'$; {\it i.e.}, 
$pp\to WZ'\to Wjj$. 
A cross section of $8.1$ pb is expected in order to explain the CDF anomaly, but is excluded using the data sample of 5 fb$^{-1}$ at 7 TeV~\cite{Chatrchyan:2012jra}. 
In all our scenarios, the $WZ'$ production cross section is smaller than $1$ pb as shown in Fig.~\ref{cross_ZpV}.  
Consequently, the $Z'$ boson in our scenarios is not constrained by current $Wb\bar{b}/Z b\bar{b}$ and $Wjj/Zjj$ events.
As we will show in the next section, the $pp \to Z' \gamma$ process serves more useful as it gives a larger cross section.

\subsection{The constraints from UA2}

\begin{figure}[t] 
\begin{center}
\includegraphics[width=120mm]{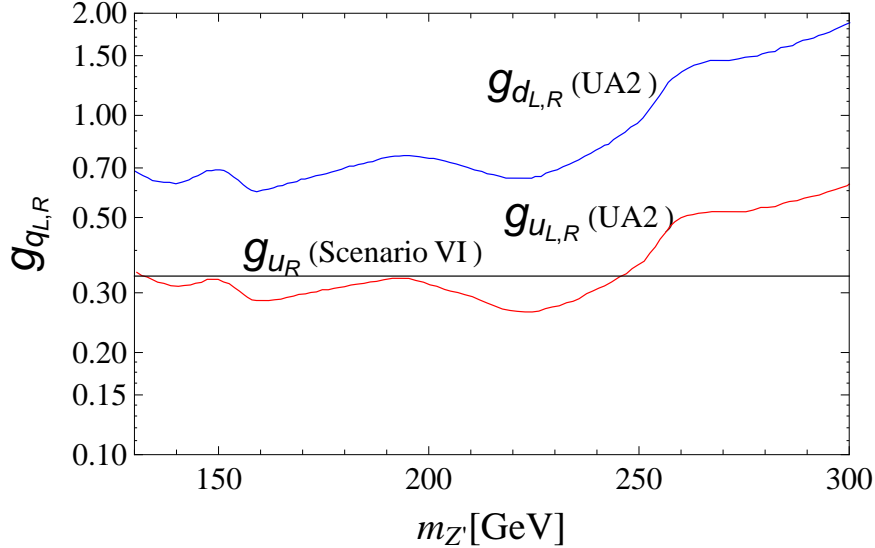} 
\caption{
Constraints on the $Z'{\bar q}q$ couplings (blue curve for the down-type quarks and red curve for the up-type quarks) taken from Ref.~\cite{Buckley} compared with $g_{u_R}$ in Scenario-VI (black horizontal line).
\label{SigZpAbb}}
\end{center} 
\end{figure}

The dijet invariant mass spectrum of $p \bar{p} \to 2$ jets had been measured in the UA2 experiment at CERN 
with the data sample of 10.9 pb$^{-1}$, in search of any extra heavy vector bosons that decay into two jets~\cite{UA2}. 
The fact that no excess had been observed can be converted into a constraint on the $Z'$ couplings 
with quarks in the mass range $130<m_{Z'}<300$ GeV.  
In Ref.~\cite{Buckley}, 
constraints on the chiral couplings in
\begin{equation}
g_{q_{L,R}} \bar{q} \gamma^\mu \frac{1 \mp \gamma_5}2 q Z'_\mu,~~\text{with}~~q=u,d, \label{Eq:qqZp}
\end{equation}
had been given in Fig.~1 by comparing the cross section for the $p\bar{p} \to  Z' \to  jj$ process with that of the upper limit obtained from the UA2 experiment at the center-of-mass energy of 630 GeV. 
The chiral couplings expressed in Eq.~(\ref{Eq:qqZp}) are related to the variables in Eq.~(\ref{Zp_int}) by 
\begin{align}
\label{couplingRelation}
g_{q_{L,R}}=g_{Z'}(v_q\pm a_q),
\end{align}
where the values of $v_q$ and $a_q$ are given by Eq.~(\ref{va}) and Table~\ref{charges} in the Appendix.
In Fig.~\ref{SigZpAbb}, we show the upper limits for $g_{u_{R,L}}$ and $g_{d_{R,L}}$ extracted from Fig.~1 of Ref.~\cite{Buckley}, 
in comparison with $g_{u_R}$ in Scenario-VI indicated by the horizontal line.  
The coupling constant $g_{u_L}$ in Scenario-VI and that including all the other couplings 
in the other scenarios are well below these constraints and thus omitted.
We therefore obtain the excluded region of $m_{Z'} \lesssim 250$ GeV in Scenario-VI.

\subsection{Summary of constraints on $m_{Z'}$}

\begin{table}[t]
\begin{center}
\begin{tabular}{l c c c c c c} \hline\hline
 & ~Scenario-I~ & ~Scenario-III~ & ~Scenario-III'~ & ~Scenario-V~ & ~Scenario-V'~ & ~Scenario-VI \\\hline
$pp\to t\bar{t}$~ & 1000 GeV & - & - & 800 GeV & 850 GeV & 1900 GeV \\ 
dijet & -& - & - & - & - & 500 GeV\\ 
UA2   & -& - & - & - & - & 250 GeV\\\hline\hline
\end{tabular}
\caption{Lower bounds on $m_{Z'}$ at 95\% CL for all the scenarios by existing experiments. 
The bounds from $pp \to  t \bar{t}$ are only valid for $m_{Z'} > 2 m_t$. }
\label{const_summary}
\end{center}
\end{table}

Here we summarize the constraints on $m_{Z'}$ discussed in this section.
The lower bounds on $m_{Z'}$ at 95\% CL for all the scenarios are listed in Table~\ref{const_summary}. 
The constraint from the $pp\to t\bar{t}$ process is the strongest among all. 
However, this constraint is valid only when the $Z'\to t\bar{t}$ decay is kinematically allowed. 
On the other hand, the constraints from the dijet events and the UA2 constraint can be applied to the lighter $Z'$ case; namely, $m_{Z'}<2m_t$. 
Except for Scenario-VI, all other scenarios are not restricted by them mainly because of the small $Z'$ charges for up-type quarks. 
We also find that it is currently difficult to extract constraints on the $Z'$ mass from the $Z b \bar{b}$, $W b \bar{b}$, $Z j j $ and $W j j$ events 
because of the small cross sections of $Z Z'$ and $W Z'$ production.
In the next section, we study the light $Z'$ case using the $t$-channel $pp\to Z' \gamma$ process.

\section{Photon associated production of $Z'$ \label{sec:simulation}}

In this section, we propose to use the $t$-channel process $pp \to  Z' \gamma \to  b \bar{b} \gamma$ to search for a relatively light $Z'$ with $m_{Z'} \lesssim 350$ GeV, where the $Z' \to  t\bar{t}$ decay is kinematically forbidden.  Since no experimental data exist at the time of writing, we present a simulation study here.
With b-tagging, the $Z' \to  b \bar{b}$ decay is expected to have higher sensitivity than the $Z' \to  q\bar{q}$ decays, where $q$ refers to quarks in the first and second generations.
For the backgrounds, we include the SM irreducible background $p p \to  b \bar{b} \gamma$ and the $p p \to  q\bar{q} \gamma$ process with mis-tagging of the $b$ quarks.

\begin{table}[t]
\begin{center}
\begin{tabular}{lcccc} \hline \hline
 & ~$\gamma Z'(\to  b\bar{b}) $~ & $\gamma b \bar{b}$ ~ & $\gamma j j$~ & ~~$\mathcal{S}$~~ \\ 
\hline
 Basic cuts in Eq.~(\ref{basic}) & 0.256& 21.8 & 2952 & 0.66 \\ 
 Double b-tagging & 0.0269 & 2.39 & 3.60 & 1.54  \\ 
 $M_{b\bar{b}}$ cut in Eq.~(\ref{Mbb}) & 0.015 & 0.449 & 0.541 &  2.14
\\ \hline \hline
\end{tabular}
\caption{Cross sections of signal and background processes in Scenario-I in units of pb, assuming $m_{Z'}=200$ GeV and $\sqrt{s}=8$ TeV as an example.  To calculate the significance $\mathcal{S}$, we take an integrated luminosity of 19.6 fb$^{-1}$. 
The basic cuts Eq.~(\ref{basic}) are imposed, and the double b-tagging is applied
after {\tt PGS} detector simulations.
 \label{events}}
\end{center}
\end{table}

In our analysis, we generate signal and background events using {\tt MADGRAPH/MADEVENT}~\cite{Ref:MG} and the {\tt CTEQ6L} PDF's for the collision energies $\sqrt{s}=$ 8 TeV and 14 TeV.
The generated events are passed onto {\tt PYTHIA}~\cite{Ref:Pythia} through the {\tt PYTHIA-PGS} package to include initial-state radiation, final-state radiation and hadronization effects.  
The detector level simulation is then carried out by {\tt PGS}~\cite{Ref:PGS}.
For the generated events, we first apply the following basic kinematical cuts:
\begin{align}
& p_T({\rm jets}) > 40~{\rm GeV}, \quad
p_T(\gamma) > 10~{\rm GeV}, \notag\\
&|\eta({\rm jets})| < 2.4, \quad
|\Delta \eta_{jj}| <2.4,  \label{basic} \\
&90~{\rm GeV} < M_{jj} < 360~{\rm GeV}, \nonumber
\end{align} 
where the jets include b-jets, and $|\Delta \eta_{jj}|$ is the rapidity difference of the two jets.  
The $M_{jj}$ cut restricts ourselves to the mass regime $100$ GeV $\alt m_{Z'} \alt 360$ GeV.
Moreover, we impose double b-tagging after the {\tt PGS} detector simulation to reduce the background events.
Afterwards, we further take a cut on the invariant mass of the two b-jets $M_{b \bar{b}}$: 
\begin{align}
 & m_{Z'}(1-0.2) < M_{b\bar{b}} < m_{Z'}+10~{\rm GeV} \label{Mbb} 
\end{align}
for each $m_{Z'}$ value\footnote{
Because the shape of the $b\bar{b}$ invariant mass distribution is asymmetric around $M_{b\bar{b}}=m_{Z'}$, 
we use different upper and lower cut limits in Eq.~(\ref{Mbb}). 
The asymmetric distribution is due to the fact that signal events tend to shift to lower $M_{bb}$ at the detector level, mainly as a result of soft radiation from the b-jets.}. 
We then calculate the signal significance defined by~\cite{Ref:significance}
\begin{equation}
\mathcal{S}=\sqrt{2[(s+b)\ln (1+s/b)-s]}, 
\label{significance}
\end{equation}
where $s$ and $b$ are the numbers of signal and background events, respectively.  
In Table.~\ref{events}, we show as an example how the cuts affect the number of events in Scenario-I, assuming $m_{Z'}=200$ GeV and $\sqrt{s}=8$ TeV.
The $\gamma jj$ background is significantly reduced by about three orders of magnitude by double b-tagging while the signal and irreducible background are down by one order of magnitude, thereby enhancing the significance by more than a factor of two.  We also observe that the $M_{b \bar{b}}$ cut is effective in reducing not only the $\gamma jj$ background but also the $b\bar{b}\gamma$ background. 

\begin{figure}[t] 
\begin{center}
\includegraphics[width=83mm]{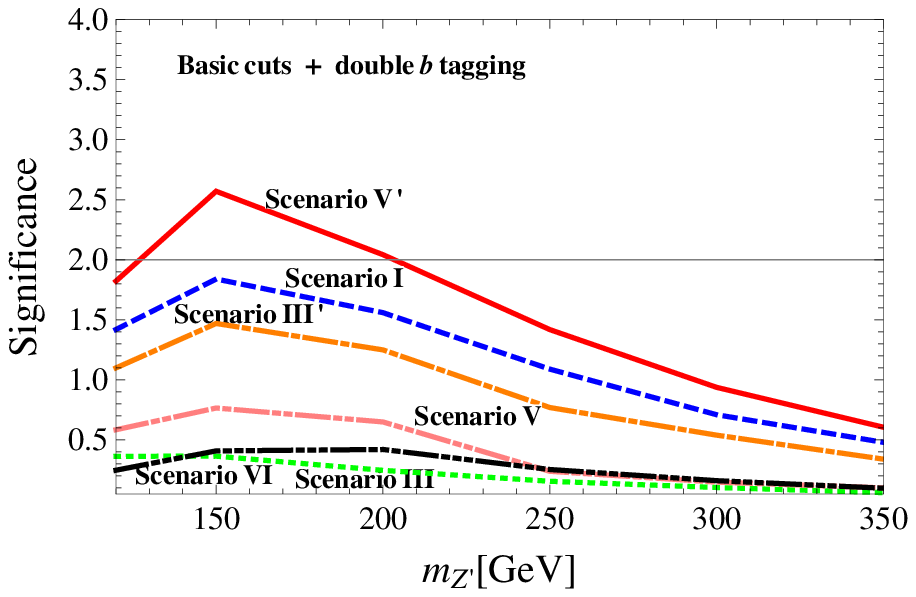}\hspace{-5mm}
\includegraphics[width=83mm]{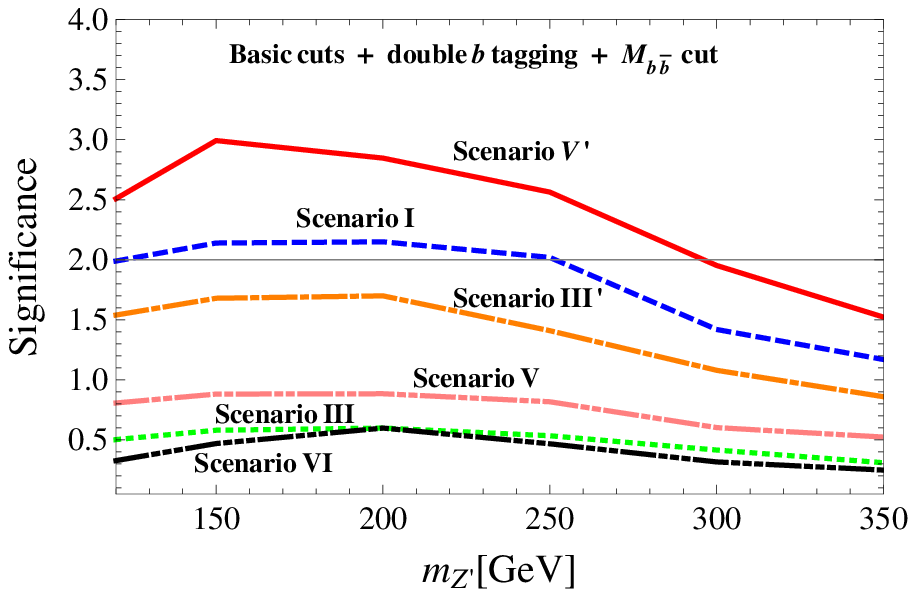}  
\caption{
Significance for the $pp \to  Z' \gamma \to  b \bar{b} \gamma$ process at the LHC with $\sqrt{s}=8$ TeV and an integrated luminosity of $19.6$fb$^{-1}$. 
The left panel shows the result after imposing the basic cuts and double b-tagging, and the right panel that after taking the $M_{b\bar{b}}$ cut.  
\label{SigZpABasic}}
\end{center} 
\end{figure}

In Fig.~\ref{SigZpABasic}, 
we show the signal significance 
for the six scenarios as a function of $m_{Z'}$, assuming $\sqrt{s}=8$ TeV and an integrated luminosity of 19.6 fb$^{-1}$. 
The left panel shows the result after imposing the basic cuts and double b-tagging, and 
the right panel that after taking the $M_{b\bar{b}}$ cut. 
Assuming the data are consistent with SM prediction, 
we find that the mass range of $130~{\rm GeV} \lesssim m_{Z'} \lesssim 200$ GeV can be excluded at 95\% CL for Scenario-V' using the basic cuts and double b-tagging.
After further imposing the $M_{b\bar{b}}$ cut, 
the mass range of $120~{\rm GeV} \lesssim m_{Z'} \lesssim 290~(240)$ GeV can be excluded for Scenario-V' (Scenario-I) at 95\% CL. 
The other scenarios will be less constrained by this analysis.
The hierarchy in the significances of the different scenarios depends on the product of the cross section of $pp \to  Z' \gamma$, shown in Fig.~\ref{cross_ZpV}, and the branching fraction of $Z' \to  b \bar{b}$, given in Table.~\ref{br_tab} and Fig.~\ref{contour}. 
%

\begin{figure}[t] 
\begin{center}
\includegraphics[width=83mm]{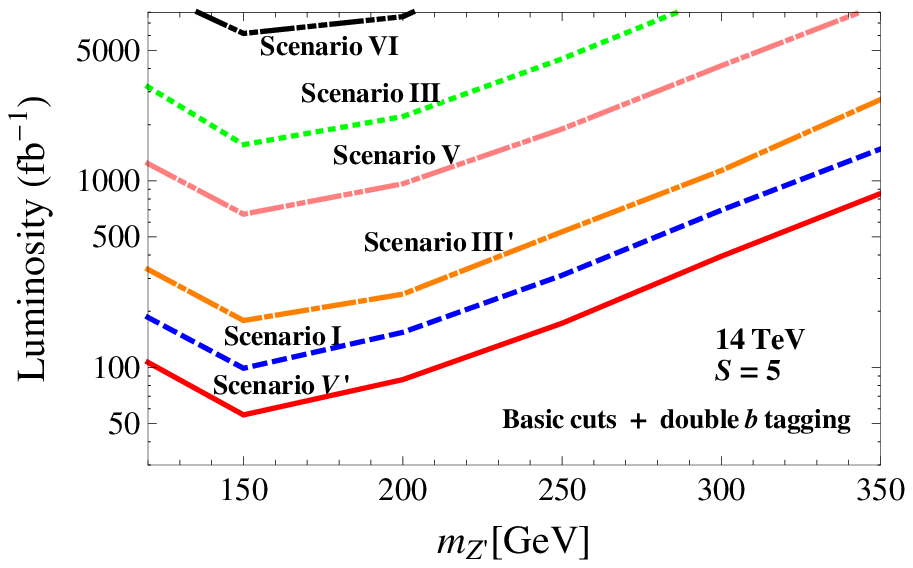} \hspace{-5mm}
\includegraphics[width=83mm]{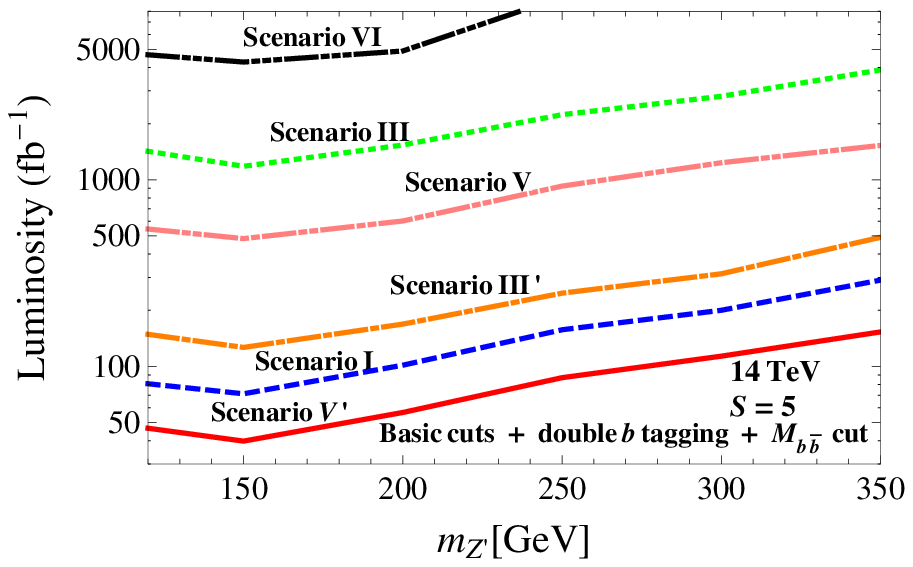} 
\caption{
Required integrated luminosity for 5-sigma discovery ($\mathcal{S}=5$) from the $pp \to  Z' \gamma \to  b \bar{b} \gamma$ process at the LHC with $\sqrt{s}=14$ TeV. 
The left panel shows the result after imposing the basic cuts and double b-tagging, and the right panel that after taking the $M_{b\bar{b}}$ cut.
}
\label{SigZpABasic14S5}
\end{center} 
\end{figure}

To study the discovery reach for the 14-TeV LHC, we compute the required integrated luminosity to reach $S=5$ as a function of $m_{Z'}$. 
The left panel in Fig.~\ref{SigZpABasic14S5} shows the result after imposing the basic cuts and double b-tagging. 
The right panel shows the result after further taking the $M_{b\bar{b}}$ cut.   
We thus find that a 5-sigma discovery can be obtained for $m_{Z'} \lesssim 200$ GeV in Scenario-V' with an integrated luminosity of $100$ fb$^{-1}$  
by applying only the basic cuts and double b-tagging.
With further the $M_{b\bar{b}}$ cut, the reach can be extended to $m_{Z'}\lesssim 290$ GeV in Scenario-V' and $m_{Z'} \lesssim 190$ GeV in Scenario-I.
With an integrated luminosity of 500 fb$^{-1}$ and all cuts mentioned above, we can cover the entire mass range in the plots for Scenarios-V', -I, and -III'.

Finally, we would like to comment on the search for the leptophobic $Z'$ boson at the international linear collider (ILC)~\footnote{
Recently, phenomenology of $Z'$ at the ILC has been discussed in Ref.~\cite{Zp_ILC} in a model-independent way using the $e^+e^-\to \mu^+\mu^-/\tau^+\tau^-$ processes. }. 
Even though the ILC is an electron-positron collider where the initial leptons do not couple to the leptophobic $Z'$ boson, it can still be produced in association with the quark pair production, {\it i.e.}, 
$e^+e^-\to \gamma^*/Z^* \to q\bar{q}Z'$.  
There are several designs of the ILC collision energy, namely, 250 GeV, 350 GeV and 500 GeV.  Therefore, 
the $Z'$ boson with a mass smaller than these energies can be produced in association with the light quark pair\footnote{
Although one can consider the $e^+e^-\to t\bar{t}Z'$ process, it suffers from a kinematical disadvantage. }.  
In particular, it would be interesting to analyze the case with $m_{Z'}<2m_t$ because it is difficult to probe using the current experimental data, as we have discussed in the previous section. 
This will lead to the signal of four hard jets with the invariant mass of two sitting at the $Z'$ mass.  
For example, the cross section of $e^+e^- \to q\bar{q}Z'$ with $m_{Z'} = 150$ GeV at a 250-GeV ILC is estimated to be $0.1$ fb.
If such a $Z'$ is discovered in the 14-TeV LHC, one may use the above-mentioned process to study detailed properties of the new boson.
Even in the case where such a boson is not found at the LHC and the exclusion limit has not reached the 95\% CL,  
one can use the this process as a discovery channel.  In any case, it is important to prepare the simulation study for the $Z'$ boson at the ILC.  
A detailed analysis for leptophobic $Z'$ searches at the ILC will be presented in a separate work~\cite{Zp_ILC2}.

\section{Conclusions\label{sec:conclusions}}

We have studied the phenomenology of a leptophobic $Z'$ boson that can be derived from the $E_6$ GUT model, as a result of kinetic mixing between $U(1)_Y$ and the extra $U(1)$ symmetries. 
Due to different embedding schemes for the matter fields in the $E_6$ fundamental representation, there are eight possible scenarios with a leptophobic $Z'$, differing in the $Z'$ charges for the quarks and the exotic fermion.  Three of the scenarios have the same $Z'$ charge ratios.  Therefore, the production cross sections of these scenarios can be related to one another by simple scaling.  This reduces the number of distinct scenarios to six.

We have taken into account current experimental data to constrain the $Z'$ mass, including $pp\to t\bar{t}$, the dijet events, $W/Z$ plus dijet events and the UA2 data. 
When the top pair decay of $Z'$ is kinematically allowed, the strongest bound on the $Z'$ mass comes from the $pp\to t\bar{t}$ process.   
In this case, the lower bounds on $m_{Z'}$ at 95\% CL are about 1 TeV, $0.8$ TeV, $0.85$ TeV and $1.9$ TeV in Scenario-I, Scenario-V', Scenario-V and Scenario-VI, respectively. 
However, this channel is not effective when $m_{Z'} < 2m_t$. 
On the other hand, only Scenario-VI is constrained by the dijet and UA2 data, from which the lower limits are given as $500$ GeV and $250$ GeV, respectively.
We also found that it is difficult to obtain constraints on $Z'$ mass from $W/Z$ plus dijet events due to the small cross sections of $ZZ'$ and $WZ'$ production.

We have proposed to use the photon associated production of the $Z'$ boson followed by the decay into a pair of bottom quarks, {\it i.e.}, $pp\to \gamma Z'\to \gamma b\bar{b}$ 
to explore the constraints in the lower mass regime, particularly for scenarios other than Scenario-VI.  Specifying the decay of $Z'$ into a pair of $b$ quarks helps reducing background events significantly.
We have performed a detailed simulation of signal and irreducible background events, and searched for appropriate kinematical cuts to increase signal significance.
We have found that Scenario-I (usually called the standard $E_6$) with $m_{Z'}\lesssim 250$ GeV can be excluded 
at 95\% CL after imposing all the cuts on the current LHC 19.6 fb$^{-1}$ data at 8 TeV.  A similar bound of $m_{Z'}\lesssim 300$ GeV has also been obtained for Scenario-V'.
Assuming an integrated luminosity of 100~fb$^{-1}$ for the 14-TeV LHC, 
a 5-sigma discovery can be reached for $m_{Z'}\lesssim 290$ GeV in Scenario-V' and $m_{Z'} \lesssim 190$ GeV in Scenario-I.

\section*{Acknowledgments}

C.-W.~C would like to thank D.~Choudhury and N.~Gaur for useful discussions during the very early stage of this project.  This research was supported in part by the National Science Council of R.~O.~C. under Grant 
Nos.~NSC-100-2628-M-008-003-MY4, NSC-101-2811-M-008-014 and NSC-102-2811-M-006-035.

\section*{APPENDIX: Review of $E_6$ GUT model}

We present a brief review of scenarios in the $E_6$ GUT model that predict a leptophobic $Z'$ boson, with its detailed derivations and fermion interactions given in Ref.~\cite{Rizzo}. 
The symmetry breaking of $E_6$ follows the pattern of
\begin{eqnarray}
 E_6 \to SO(10) \times U(1)_\psi \to SU(5) \times U(1)_\chi \times U(1)_\psi \to G_{\text{SM}} \times  U(1)_{Q'}. 
\end{eqnarray}
The $U(1)_{Q'}$ symmetry is obtained as a linear combination of $U(1)_\psi$ and $U(1)_\chi$, with the corresponding charge expressed as
\begin{equation}
Q' = Q_\psi \cos \theta - Q_\chi \sin \theta, \label{comb_Q}
\end{equation}
where $Q_\psi$ and $Q_\chi$ are the charges under $U(1)_\psi$ and $U(1)_\chi$, respectively. 

The most general kinetic terms, including kinetic mixing, for the gauge fields and interaction terms for a fermion $\psi$ are given by
\begin{align}
\mathcal{L}_{\text{kin}} &= -\frac{1}{4} W^a_{\mu \nu} W^{a \mu \nu} 
- \frac{1}{4}(\tilde{B}_{\mu\nu},\tilde{Z}_{\mu\nu}')
\left(
\begin{array}{cc}
1 & \sin\chi\\
\sin\chi & 1
\end{array}
\right)
\left(
\begin{array}{c}
\tilde{B}^{\mu\nu}\\
\tilde{Z}^{'\mu\nu}
\end{array}\right) ~,  \label{kin} \\
\mathcal{L}_{\text{int}} &= - \bar{\psi} \gamma^\mu ( gT^a W^a_\mu + g' Y \tilde{B}_\mu + \tilde{g}_{Z'} Q' \tilde{Z'}_\mu) \psi ~, 
\end{align}
where $g$ and $g'$ are the SM $SU(2)_L$ and $U(1)_Y$ gauge couplings, and $Y$ is the SM $U(1)_Y$ hypercharge\footnote{
We do not pull out the factors of $\sqrt{3/5}$ and $\sqrt{5/3}$ for $Y$ and $g'$, respectively. }. 
We note in passing that the kinetic mixing can be obtained at one-loop level with the matter contents of non-zero $Q'$ and $Y$ charges running inside the loop~\cite{Holdom:1985ag,Dienes:1996zr}. 
In this case, the kinetic mixing $\sin \chi$ would be proportional to the factor $g' \tilde{g}_{Z'}/(24 \pi^2) \sum Q' Y \ln (|q|^2/M_{GUT}^2)$, where the sum is taken over the matter contents inside the loop, and 
$|q|$ is the electroweak scale and $M_{GUT}$ is the GUT scale.
Then it would be possible to obtain the kinetic mixing of $O(0.1)$ to $O(1)$ required to realize leptophobia.
Through a non-unitary transformation, 
\begin{align}
\left(
\begin{array}{c}
\tilde{B}_\mu\\
\tilde{Z'}_\mu
\end{array}\right)=
\left(
\begin{array}{cc}
1 & -\tan\chi\\
0 & \sec\chi
\end{array}\right)
\left(
\begin{array}{c}
B_\mu\\
Z'_\mu
\end{array}\right) ~, 
\end{align}
the gauge fields $\tilde{Z}'_\mu$ and $\tilde{B}_\mu$ are diagonalized to the fields $Z'_\mu$ and $B_\mu$ of mass eigenstates. 
The interaction terms are then rewritten as
\begin{equation}
\mathcal{L}_{\text{int}} = - \bar{\psi} \gamma^\mu ( g T^a W^a_\mu + g' Y B_\mu 
+ g_{Z'} \bar{Q} Z'_\mu )\psi ~, 
\end{equation}
where $g_{Z'}\equiv \tilde{g}_{Z'}/\cos\chi$, and the $Z'$ charge $\bar{Q}$ for a fermion field $f$ is 
\begin{align}
\bar{Q}(f)\equiv Q'(f) + \sqrt{\frac{3}{5}}\delta Y ~,~~\text{with}~~\delta\equiv-\sqrt{\frac{5}{3}}\frac{g'}{g_{Z'}}\tan\chi ~.  
\label{qbar}
\end{align}
As is evident from Eq.~(\ref{qbar}), all $\bar{Q}(f)$ charges are determined by two unknown parameters; {\it i.e.}, $\theta$ and $\delta$. 
This implies that once we fix two of $\bar{Q}(f)$ charges, all the other $\bar{Q}(f)$ charges are uniquely determined as well. 
We utilize this feature to find scenarios with a leptophobic $Z'$; that is,
\begin{align}
 \bar{Q}(L) =  Q'(L)-\frac{1}{2}\delta =0 ~, \quad
 \bar{Q}(e^c) = Q'(e^c)+\delta =0 ~. 
\end{align}
These two equations are solved to render 
\begin{align}
\tan\theta =\frac{2Q_\psi(L)+Q_\psi(e^c)}{2Q_\chi(L)+Q_\chi(e^c)},\quad \delta = -Q'(e^c)~(=2Q'(L)) ~. 
\end{align}

There are then six ways to embed the SM fermions along with exotic fermion denoted by $h$ into the {\bf 27} representation of $E_6$. 
In Table~\ref{charges}, $U(1)$ charges of the fermions are listed for the six scenarios, following the convention of Ref.~\cite{London}. 
In this table, only $h$ is a non-SM fermion whose SM gauge quantum numbers are the same as those of $d$. 
Thus, one can interchange the $U(1)_\psi$ and $U(1)_\chi$ charges of $d$ with those of $h$. 
After the interchange, the $Q'$ and $\bar{Q}$ charges of $d$ and $h$ are different from the original ones only in Scenario-III and Scenario-V.  We denote the two new scenarios by Scenario-III' and Scenario-V', respectively.

\begin{table}[t]
\begin{center}
\begin{tabular}{l|l|l|l|l||l|l|l|l||l|l|l|l} \hline \hline
&\multicolumn{4}{c||}{Scenario-I  }&\multicolumn{4}{c||}{Scenario-II } &\multicolumn{4}{c}{Scenario-III (Scenario-III')}\\ \hline 
&\multicolumn{4}{c||}{$\tan\theta=\sqrt{3/5}$, $\delta=-1/3$}&\multicolumn{4}{c||}{$\tan\theta=\sqrt{15}$, $\delta=-\sqrt{10}/3$} &\multicolumn{4}{c}{$\tan\theta=\sqrt{5/3}$, $\delta=-\sqrt{5/12}$}\\ \hline 
 & $2 \sqrt{6} Q_\psi$ & $2 \sqrt{10} Q_\chi$  & $ \sqrt{15}Q'$ & $ \sqrt{15} \bar{Q}$
 & $2 \sqrt{6} Q_\psi$ & $2 \sqrt{10} Q_\chi$  & $ \sqrt{6}Q'$ & $ \sqrt{6}\bar{Q}$
 & $2 \sqrt{6} Q_\psi$ & $2 \sqrt{10} Q_\chi$  & $Q'$ & $\bar{Q}$ \\ \hline 
 $Q$  & $1$ & $-1$ & $1$ & $5/6$       & $1$ & $-1$ & $1/2$   & $1/6$      & $1$ & $-1$ & $1/4$ & $1/6$ \\
 $u^c$&$1$  & $-1$ & $1$ & $5/3$       & $1$ & $3$  & $-1$  & $1/3$        & $1$ & $3$  & $-1/4$ & $1/12$ \\
 $d^c$&$1$  & $3$  & $-1/2$& $-5/6$    & $1$ & $-1$ & $1/2$   & $-1/6$     & $1$~$(1)$ & $-1$~$(3)$ & $1/4$~$(-1/4)$ & $1/12$~$(-5/12)$ \\
 $h^c$&$-2$  & $-2$  & $-1/2$& $-5/6$  & $-2$ & $-2$ & $1/2$ & $-1/6$      & $1$~$(1)$ & $3$~$(-1)$ & $-1/4$~$(1/4)$ & $-5/12$~$(1/12)$  \\
 $L$  &$1$  & $3$  & $-1/2$& $0$       & $1$ & $3$  & $-1$  & $0$          & $1$ & $3$  & $-1/4$ & $0$\\
 $e^c$&$1$  & $-1$ & $1$ & $0$         & $1$ & $-5$& $2$    & $0$          & $4$ & $0$& $1/2$  & $0$\\
 \hline \hline 
\end{tabular}\\ \vspace{5mm}
\begin{tabular}{l|l|l|l|l||l|l|l|l||l|l|l|l} \hline \hline
&\multicolumn{4}{c||}{Scenario-IV}&\multicolumn{4}{c||}{Scenario-V (Scenario-V') } &\multicolumn{4}{c}{Scenario-VI}\\ \hline 
&\multicolumn{4}{c||}{$\tan\theta=\sqrt{3/5}$, $\delta=-1/3$}&\multicolumn{4}{c||}{$\tan\theta=\sqrt{5/27}$, $\delta=-\sqrt{5/12}$} &\multicolumn{4}{c}{$\tan\theta=0$, $\delta=-\sqrt{10}/3$}\\ \hline 
 & $2 \sqrt{6} Q_\psi$ & $2 \sqrt{10} Q_\chi$  & $ \sqrt{15}Q'$ & $ \sqrt{15} \bar{Q}$
 & $2 \sqrt{6} Q_\psi$ & $2 \sqrt{10} Q_\chi$  & $ Q'$ & $ \bar{Q}$
 & $2 \sqrt{6} Q_\psi$ & $2 \sqrt{10} Q_\chi$  & $ \sqrt{6}Q'$     & $ \sqrt{6} \bar{Q}$ \\ \hline 
 $Q$    & $1$  & $-1$   & $1$ & $5/6$       & $1$  & $-1$ & $1/4$   & $1/6$    & $1$  & $-1$  & $1/2$  & $1/6$ \\
 $u^c$  &$1$   & $-1$   & $1$ & $5/3$       & $1$  & $3$  & $0$     & $1/3$    & $1$  & $3$   &$1/2$   & $11/6$  \\
 $d^c$  &$1$   & $3$    & $-1/2$& $-5/6$    & $1$~$(-2)$  & $-1$~$(-2)$ & $1/4$~$(-1/4)$   & $1/12$~$(-5/12)$   & $1$  & $-1$  & $1/2$  & $-1/6$ \\
 $h^c$  &$-2$  & $-2$   & $-1/2$  & $-5/6$  & $-2$~$(1)$ & $-2$~$(-1)$ & $-1/4$~$(1/4)$  & $-5/12$~$(1/12)$  & $1$  & $3$   & $1/2$  & $-1/6$ \\
 $L$    &$-2$  & $-2$   & $-1/2$& $0$       & $-2$ & $-2$ & $-1/4$  & $0$      & $-2$ & $-2$  & $-1$   & $0$\\
 $e^c$  &$1$   & $-1$   & $1$ & $0$         & $1$  & $-5$ & $1/2$   & $0$      & $4$  & $0$   &  $2$   & $0$ \\
\hline \hline 
\end{tabular}
\caption{$U(1)$ charges of six leptophobic scenarios in the $E_6$ GUT model.  Fields with a superscript $c$ denote the corresponding charge-conjugated fields. }
\label{charges}
\end{center}
\end{table}

\end{document}